\documentclass[apjl]{emulateapj}
\usepackage{apjfonts}
\usepackage{CJK}

\newcommand{\wise}{{\textit{WISE}}}

\shorttitle{Merger fraction of Hot DOGs}
\shortauthors{Fan et al.}

\begin{document}

\begin{CJK*}{UTF8}{gbsn}

\newcommand{\ser}{S\'ersic~}
\newcommand{\gf}{{\tt GALFIT}~}

\title{The most-luminous heavily-obscured quasars have a high merger fraction: morphological study of \wise-selected hot dust-obscured galaxies}

\author{Lulu Fan (范璐璐)\altaffilmark{1,\dag}, Yunkun Han (韩云坤)\altaffilmark{2,\ddag}, Guanwen Fang (方官文) \altaffilmark{3}, Ying Gao (高颖)\altaffilmark{1}, Dandan Zhang (张丹丹) \altaffilmark{1}, Xiaoming Jiang (蒋效铭) \altaffilmark{1}, Qiaoqian Wu (伍巧倩) \altaffilmark{1},  Jun Yang (杨俊)\altaffilmark{1} and Zhao Li (李钊) \altaffilmark{1}}
\altaffiltext{1}{Shandong Provincial Key Lab of Optical Astronomy and Solar-Terrestrial Environment, Institute of Space Science, Shandong University,Weihai, 264209, China}
\altaffiltext{2}{Yunnan Observatories, Chinese Academy of Sciences, Kunming, 650011, China}
\altaffiltext{3}{Institute for Astronomy and History of Science and Technology, Dali University, Dali 671003, China}
\altaffiltext{\dag}{llfan@sdu.edu.cn}
\altaffiltext{\ddag}{hanyk@ynao.ac.cn}

\begin{abstract}
Previous studies have shown that \wise-selected hyperluminous, hot dust-obscured galaxies (Hot DOGs) are powered by highly dust-obscured, possibly Compton-thick AGNs. High obscuration provides us a good chance to study the host morphology of the most luminous AGNs directly. We analyze the host morphology of 18 Hot DOGs at $z\sim3$ using Hubble Space Telescope/WFC3 imaging. We find that Hot DOGs have a high merger fraction ($62\pm 14 \%$). By fitting the surface brightness profiles, we find that the distribution of \ser indices in our Hot DOG sample peaks around 2, which suggests that most of Hot DOGs have transforming morphologies. We also derive the AGN bolometric luminosity ($\sim10^{14}L_\odot$) of our Hot DOG sample by using IR SEDs decomposition. The derived merger fraction and AGN bolometric luminosity relation is well consistent with the variability-based model prediction \citep{hickox2014}. Both the high merger fraction in IR-luminous AGN sample and relatively low merger fraction in UV/optical-selected, unobscured AGN sample can be expected in the merger-driven evolutionary model. Finally, we conclude that Hot DOGs are merger-driven and may represent a transit phase during the evolution of massive galaxies, transforming from the dusty starburst dominated phase to the unobscured QSO phase.
\end{abstract}

\keywords{galaxies: active --- galaxies: evolution --- galaxies: high-redshift --- galaxies: interactions}

\section{Introduction}

Both cosmic star formation rate and active galactic nucleus (AGN) density have been found to reach their peaks at $z\sim$2 \citep{reddy2008,brandt2015}. In the local Universe, a supermassive black hole (SMBH) generically exists in the center of early-type galaxies with the black hole mass tightly correlating with that of the galaxy's stellar bulge \citep{magorrian1998,ferrarese05}. The connection and co-evolution between the central SMBH and host galaxy have therefore been suggested \citep{alexander2012, kormendy2013}. In one of the most popular co-evolution scenarios, galaxy mergers have been proposed to funnel gas into the center of galaxies, leading to a central starburst and rapid growth of a SMBH \citep{hopkins2008}.

One promising approach to investigate the merger-driven co-evolution scenario is to study the merger features in AGN host galaxies. However, previous studies have produced mixed results. On one side, the most moderate-luminosity X-ray selected AGN hosts ($L_X<10^{44}$erg~s$^{-1}$) have similar disk-dominated morphologies as those of non-active galaxies, showing no significant difference in the distortion fraction, both at $z<1$ \citep{cisternas2011,villforth2014} and at $z\sim2$ \citep{schawinski2011,kocevski2012,fan2014,rosario2015}. On the other side, the high merger fraction ($\sim85\%$) has been found in a subsample of the bright ($L_{bol}>10^{46}$erg~s$^{-1}$), dust-reddened quasars \citep{urrutia2008}. This
may lead to an explanation that merger fraction is dependent on AGN bolometric luminosity \citep{treister2012}. There are also theoretical studies suggesting that galaxy mergers only trigger luminous AGN activity while other internal mechanisms may be responsible in less luminous AGNs \citep{hopkins2009,draper2012}. Therefore, it is crucial to examine the connection between the most luminous AGNs and merger fractions. However, host morphological studies of the most luminous AGNs ($L_{bol}>10^{46}$erg~s$^{-1}$) at $z\sim2$ are rare in the literature. For the luminous blue AGNs, such studies have been challenged by the bright point source, even with the careful treatment of point source substraction \citep{mechtley2015}. The present sampling in deep and narrow surveys has  biased against the luminous X-ray selected AGNs.

NASA's \textit{Wide-field Infrared Survey Explorer} (\textit{WISE}; Wright et al. 2010) all-sky survey provides an opportunity to search the most luminous galaxies at mid-infrared wavelengths. \citet{eisenhardt2012} and \citet{wu2012} discovered a new population of hyperluminous, hot dust-obscured galaxies (thereafter Hot DOGs) using a so-called "W1W2 dropout" selection criteria. Follow-up studies have revealed several key aspects of this population: (1) Spectroscopic follow-up studies show that they are mostly high-redshift objects, with redshift range from 1 to 4 \citep{wu2012,tsai2015}. (2) Most of Hot DOGs are extremely luminous with $L_{bol}>10^{13}L_\odot$ \citep{wu2014,jones2014}. (3) Using X-ray observations \citep{stern2014,piconcelli2015,assef2015b} and spectral energy distribution (SED) analysis (Assef et al. 2015a; Fan et al. 2016a), clear evidence has been shown that their luminous mid-IR emission is powered by a highly dust-obscured, possibly Compton-thick AGN. Thanks to the heavy obscuration of circumnuclear dust, the host morphology of this population is easily observable. Thus Hot DOGs are ideal objects for us to investigate the merger fraction in the most luminous AGNs.

In this letter, we examine the host morphology of 18 Hot DOGs, which have \textit{Hubble Space Telescope (HST)} Wide-Field Camera 3 (WFC3) near-IR high-resolution imaging. Our target is to provide some knowledge about the merger fraction in the most luminous AGNs.
%In Section 2, we briefly describe the sample selection, near-IR and IR data for the study. In Section 3, we estimate the merger fraction of this subsample of Hot DOGs by using visual classifications. By analyzing their IR SEDs, we derive the bolometric luminosity of each Hot DOGs in Section 4. We present the result of the merger fraction as a function of AGN bolometric luminosity in Section 5. We summarize our main results and have a brief discussion in Section 6.
Throughout this work we assume a flat ${\rm \Lambda}$CDM cosmology with $H_0 = 70$ km~s$^{-1}$, $\Omega_M = 0.3$, and $\Omega_\Lambda = 0.7$.

\section{Sample and Data}

The Hot DOGs studied here are selected from the \wise~ All-Sky Source catalog \footnote{http://wise2.ipac.caltech.edu/docs/release/allwise/} \citep{cutri2013}. In order to investigate the host morphology of Hot DOGs, we select a subsample of 18 objects (Table 1) with available \textit{HST} WFC3 imaging. We also require that they have known spectroscopic redshift in the literature \citep{wu2012,jones2014,tsai2015}  for calculating their bolometric luminosities.

\subsection{Near-IR High-resolution Imaging}

To investigate the host morphology of Hot DOGs, we use the high resolution \textit{HST} WFC3 near-IR imaging. We search our targets and retrieve the calibrated images from MAST\footnote{http://archive.stsci.edu/hst/search.php}. Observations are from four different HST proposals with ID 12488 (PI: M. Negrello), 12585 (PI: S. Petty), 12481 and 12930 (PI: C. Bridge). We list HST proposal ID for each object in Table 1. All but one (W0831+0140) use the \textit{F160W} ($H$-band) imaging. Only W0831+0140 use the \textit{F110W} imaging. The WFC3 imaging has a pixel scale of 0.13 arcsec/pixel.

\begin{table}
\centering
\caption{The sample of Hot DOGs\label{tbl-sample}}
%\begin{center}
\begin{tabular}{lcccc}
\hline
\hline
Source & R.A. & Decl. & Redshift & HST Proposal ID	\\
Name & (J2000) & (J2000) &  	 &			\\
\hline
W0149+2350  	&     01:49:46.16  &     +23:50:14.6  & 3.228	&  12481	\\
W0243+4158	&     02:43:44.18  &	 +41:58:09.1  & 2.010   &  12930	\\
W0410$-$0913  	&     04:10:10.60  &   $-$09:13:05.2  & 3.592   &  12930	\\
W0542$-$2705	&     05:42:30.90  &   $-$27:05:40.5  & 2.532	&  12585	\\		
W0757+5113  	&     07:57:25.07  &     +51:13:19.7  & 2.277   &  12930	\\
W0831+0140	&     08:31:53.26  &     +01:40:10.8  & 3.888	&  12488	\\
W0851+3148	&     08:51:24.78  &     +31:48:56.1  & 2.640	&  12930	\\
W0859+4823  	&     08:59:29.94  &     +48:23:02.3  & 3.245	&  12930	\\
W1136+4236  	&     11:36:34.31  &     +42:36:02.6  & 2.390	&  12481	\\
W1316+3512	&     13:16:28.53  &     +35:12:35.1  & 1.956	&  12585	\\
W1603+2745  	&     16:03:57.39  &     +27:45:53.3  & 2.633	&  12930	\\
W1814+3412  	&     18:14:17.30  &     +34:12:25.0  & 2.452   &  12481	\\
W1830+6504	&     18:30:13.53  &     +65:04:20.5  & 2.653 	&  12585	\\
W1835+4355  	&     18:35:33.71  &     +43:55:49.1  & 2.298	&  12585	\\
W2026+0716	&     20:26:15.27  &     +07:16:23.9  & 2.540	&  12930	\\
W2207+1939	&     22:07:43.84  &	 +19:39:40.3  & 2.022	&  12930	\\
W2246$-$0526  	&     22:46:07.57  &   $-$05:26:35.0  & 4.593	&  12930	\\
W2246$-$7143	&     22:46:12.07  &   $-$71:44:01.3  & 3.458	&  12930	\\
\hline
\end{tabular}
\end{table}

\subsection{IR data}

In order to estimate the bolometric luminosity of Hot DOGs, we construct complete mid-IR to submm/mm SEDs, including \wise~ W3 and W4, \textit{Herschel} PACS \citep{poglitsch2010}
and SPIRE \citep{griffin2010}, \textit{JCMT} SCUBA-2 850$\mu m$ and other millimeter observations when available. The \wise~ W3 and W4 photometry are taken from the ALLWISE Data Release \citep{cutri2013}. W3 and W4 flux densities and uncertainties have been converted from catalog Vega magnitude by using zero points of 29.04 and 8.284 Jy, respectively \citep{wright2010}. We retrieve the \textit{Herschel} data via \textit{Herschel} Science Archive (HSA)\footnote{http://www.cosmos.esa.int/web/herschel/science-archive}. Both PACS and SPIRE data have been reduced using the \textit{Herschel} Interactive Processing Environment (HIPE v12.1.0). Several Hot DOGs also have JCMT SCUBA-2 850$\mu$m submm \citep{jones2014}, CSO and SMA millimeter observations \citep{wu2012,wu2014}.

\section{Morphological classification}

\begin{figure*}
%\epsscale{.80}
\plotone{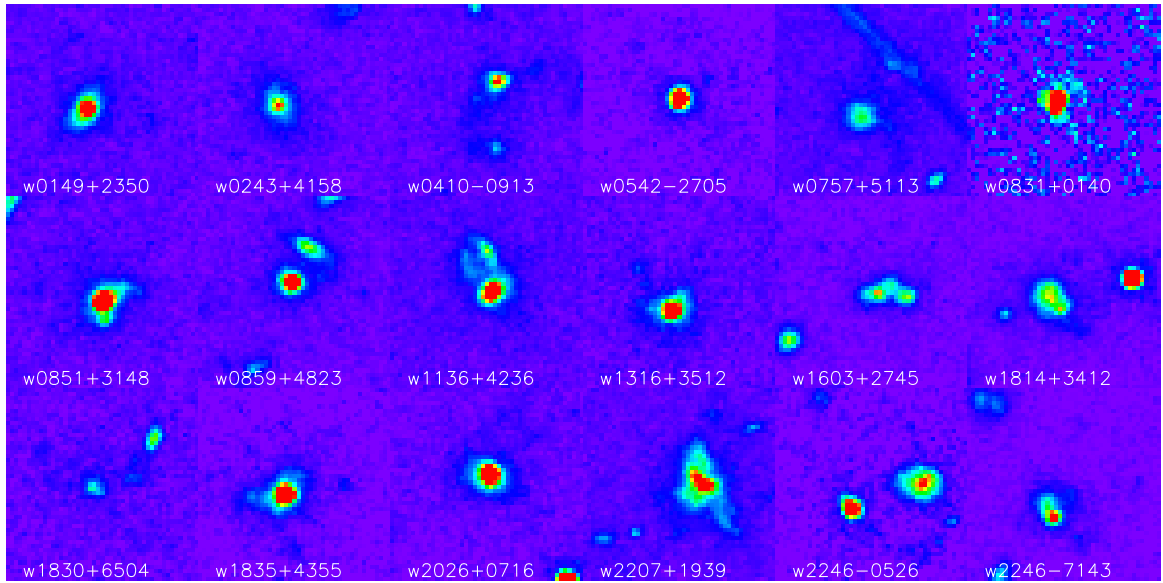}
\caption{HST WFC3 cutouts of 18 Hot DOGs which have been used for our visual classification. The size of each stamp is $6"\times 6"$.}
\end{figure*}

We investigate the merger fraction of Hot DOGs through visual classification. Different visual classification methods have been performed by different authors. \citet{cisternas2011} described the distortion degree of the galaxy morphology by defining three distortion classes: ``Dist 0" represents undisturbed and smooth galaxies, showing no interaction signatures. Class ``Dist 1" represents  galaxies with mild distortions, possibly due to minor merger or accretion. Class ``Dist 2" represents  galaxies with strong distortions, potential signs for ongoing or recent mergers. Another classification scheme \citep{kocevski2012,kocevski2015,kartaltepe2015} classified the galaxy morphological type into \textit{Disk, Spheroid, Irregular/Peculiar, Point-like} and the degree to which the galaxy is disturbed into \textit{Merger/Interaction, Disturbed/Asymmetric, Undisturbed}. For the sake of simplicity, we classify the host morphology of the Hot DOG sample into only two categories: (1) \textit{Merger/Interaction} with the merging features, such as tidal tail, shell and interaction. (2) \textit{Undisturbed} with no merging feature. Based on the WFC3 \textit{F160W} and \textit{F110W} cutouts (see Figure 1),  all 18 Hot DOGs have been visually classified by nine human classifiers independently.

In additional to visual classification, we also fit the surface brightness profiles of 18 Hot DOGs using the \gf package \citep{peng2002}. As central point source has been heavily obscured at UV/optical band, we neglect its contribution to the observed surface brightness profile. A sole \ser function has been used to model the host galaxy. We derive the structural parameter \ser indices $n$ from the best fitting result.

We also measure nonparametric morphological parameters, $Gini$ and $M_{20}$, which can be used to do the automatically morphological classification (Abraham et al. 1996; Lotz et al. 2008).

\section{AGN bolometric luminosity}

In order to estimate the bolometric luminosity of the obscured AGN in the Hot DOGs, we derive the torus emission based on the IR SEDs. The SED fitting method used here is identical to which we used in our previous work (Fan et al. 2016a). We summarize the main ideas here.

We suppose that the observed IR SEDs of Hot DOGs have been dominated by the AGN torus emission and the cold dust emission which may be related to star formation. We employ an updated version of the Bayesian SED fitting code  BayeSED (\citeauthor{Han2012a} \citeyear{Han2012a}, \citeyear{HanY2014a})\footnote{https://bitbucket.org/hanyk/bayesed/} to decompose the IR SEDs of Hot DOGs by using a new version of the CLUMPY torus model \citep{nenkova2008a}\footnote{www.clumpy.org} and a simple gray body model to represent the contribution of cold dust emission.
The gray body model is defined as:
\begin{equation}\label{equ:gb}
  S_{\lambda}=(1-e^{-(\frac{\lambda_0}{\lambda})^{\beta}}) B_\lambda(T_{dust})
\end{equation}
where $B_\lambda$ is the Planck blackbody spectrum, $T_{\rm dust}$ is dust temperature, and we use the typical value of  $\lambda_0$ = 125$\,\mu$m and adopt $\beta$=1.6.

The IR luminosities of Hot DOGs based on the best-fitting results have been derived by employing a torus plus gray body model and have been decomposed into AGN torus luminosity and cold dust luminosity (Table 2). We adopt the same bolometric correction as we used in Fan et al. (2016a): $L_{bol}=1.4\times L^{t}_{IR}$. The bolometric correction 1.4 is also consistent with the result of \citet{piconcelli2015} who found that the IR luminosity contributed to $\sim75\%$ the AGN bolometric luminosity in the Hot DOG  W1835+4355, with IR and X-ray data. The median AGN bolometric luminosity of our Hot DOG sample is $\sim 1.0\times 10^{14}L_\odot$.

\begin{table}
\centering
\caption{IR luminosities of Hot DOGs\label{tbl-lum}}
%\begin{center}
\begin{tabular}{lcc}
\hline
\hline
Source & log $L_{IR}^{t}$ & log $L_{IR}^{cd}$     \\
       & ($L_\odot$) & ($L_\odot$)  \\
\hline
  W0149+2350  &  13.89 $^{+ 0.02}_{- 0.02}$  &  13.23 $^{+ 0.05}_{- 0.05}$  \\
  W0243+4158  &  13.22 $^{+ 0.02}_{- 0.03}$  &  12.73 $^{+ 0.07}_{- 0.06}$    \\
  W0410-0913  &  14.20 $^{+ 0.02}_{- 0.02}$  &  13.70 $^{+ 0.03}_{- 0.02}$    \\
  W0542-2705  &  13.83 $^{+ 0.04}_{- 0.05}$  &  13.02 $^{+ 0.10}_{- 0.14}$   \\
  W0757+5113  &  13.42 $^{+ 0.02}_{- 0.02}$  &  12.79 $^{+ 0.03}_{- 0.03}$   \\
  W0831+0140  &  14.28 $^{+ 0.09}_{- 0.11}$  &  13.78 $^{+ 0.05}_{- 0.12}$   \\
  W0851+3148  &  13.90 $^{+ 0.03}_{- 0.04}$  &  12.87 $^{+ 0.14}_{- 0.14}$   \\
  W0859+4823  &  14.00 $^{+ 0.01}_{- 0.01}$  &  13.32 $^{+ 0.01}_{- 0.02}$   \\
  W1136+4236  &  13.61 $^{+ 0.08}_{- 0.08}$  &  13.23 $^{+ 0.04}_{- 0.04}$   \\
  W1316+3512  &  13.44 $^{+ 0.02}_{- 0.03}$  &  12.50 $^{+ 0.04}_{- 0.05}$   \\
  W1603+2745  &  13.61 $^{+ 0.02}_{- 0.02}$  &  13.24 $^{+ 0.03}_{- 0.03}$   \\
  W1814+3412  &  13.72 $^{+ 0.02}_{- 0.02}$  &  13.18 $^{+ 0.04}_{- 0.04}$   \\
  W1830+6504  &  13.50 $^{+ 0.02}_{- 0.02}$  &  12.75 $^{+ 0.05}_{- 0.06}$   \\
  W1835+4355  &  13.89 $^{+ 0.01}_{- 0.01}$  &  13.29 $^{+ 0.02}_{- 0.03}$   \\
  W2026+0716  &  13.72 $^{+ 0.05}_{- 0.05}$  &  13.09 $^{+ 0.10}_{- 0.13}$   \\
  W2207+1939  &  13.48 $^{+ 0.04}_{- 0.04}$  &  12.93 $^{+ 0.08}_{- 0.08}$   \\
  W2246-0526  &  14.46 $^{+ 0.01}_{- 0.02}$  &  13.73 $^{+ 0.04}_{- 0.05}$   \\
  W2246-7143  &  14.14 $^{+ 0.05}_{- 0.04}$  &  13.57 $^{+ 0.06}_{- 0.15}$   \\
\hline
\end{tabular}
\end{table}

\section{Results}

\begin{figure*}
\plotone{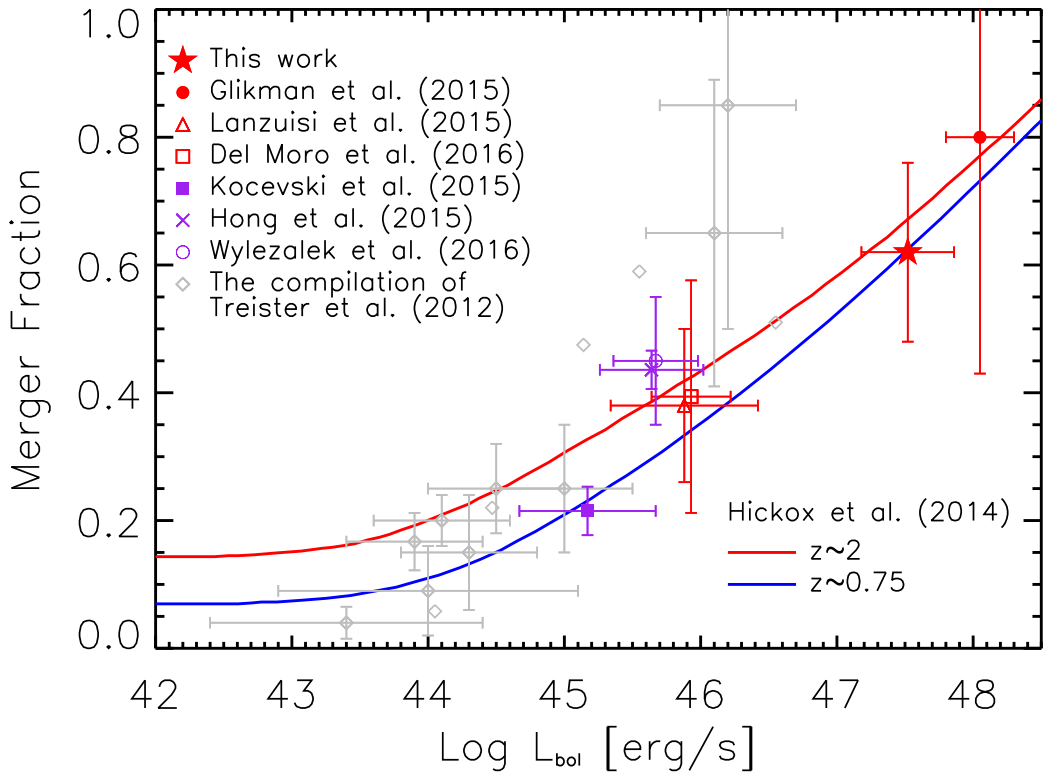}
\caption{The merger fraction as a function of AGN bolometric luminosity. Gray diamonds are taken from the compilation of \citet{treister2012}. Red and purple symbols, which represent the samples at intermediate redshift $z<1$ and at $z\sim2$, respectively, are taken from the recent works. The red asterisk represents the result in this work. The observed merger fraction and AGN bolometric luminosity relation has been compared to the variability-driven model prediction from \citet{hickox2014} realized at  $z\sim 2$ (red line) and $z\sim 0.75$ (blue line). }
\end{figure*}

We find a high merger fraction ($62\pm 14 \%$) in the Hot DOG sample, using the visual classification described in Section 3. Given the high AGN bolometric luminosity ($\sim1.0\times 10^{14}L_\odot$) in the Hot DOG sample, this finding provides a clear evidence that the most luminous AGNs have likely been related to galaxy mergers. In order to investigate the dependence of merger fraction on AGN bolometric luminosity, we compile the data in the literature and plot them in Figure 2. Gray diamonds are taken from the compilation of \citet{treister2012}. Red and purple symbols, which represent the samples at intermediate redshift $z<1$ and at $z\sim2$, respectively, are taken from the recent works \citep{glikman2015,lanzuisi2015,hong2015,kocevski2015,delmoro2016,wylezalek2016}. Our result for the Hot DOG sample has been shown as red asterisk. Among all available data, our sample is the second brightest, which has the bolometric luminosity only lower than that of the dust-reddened QSO sample in \citet{glikman2015}.

An obvious trend can be found in Figure 2 that the merger fraction increases with AGN bolometric luminosity at high luminosity, while the merger fraction shows a weak dependence on AGN bolometric luminosity for the less luminous AGNs. We compare the observed trend with the variability-driven model of \citet{hickox2014} (red and blue lines in Figure 2). Red and blue lines show the predictions of \citet{hickox2014} model at $z\sim 2$ and $z\sim 0.75$, respectively. The merger fraction in our sample agrees well with the model prediction at $z\sim2$. In other samples at $z\sim2$ of mid-IR luminous QSOs (red square, Del Moro et al. 2016), compton-thick AGNs (red triangle, Lanzuisi et al. 2015) and dust-reddened QSOs (red point, Glikman et al. 2015), the merger fractions also show a good agreement with the model prediction. As suggested by \citet{glikman2015}, galaxy merger, especially major merger, may play the most significant role in triggering the most luminous AGNs at $z\sim2$.

In Figure 3, we show the distribution of \ser indices in our Hot DOG sample. The distribution peaks at $n=2.1$, ranging from disk-dominated ($n<1.5$), disk with a prominent bulge component (1.5$\le$n$\le$3.0) to bulge-dominated ($n>3.0$). The fractions of disk-dominated, intermediate and bulge-dominated morphologies are 22.2\%, 61.1\% and 16.7\%, respectively, in our Hot DOG sample. We remind the reader that a single \ser model won't provide a good fitting for several heavily distorted Hot DOGs.  For instance, both W0851+3148 and W2207+1939 have significantly morphological distortions (see Figure 1) and the results of a single \ser model fitting have a large value of the reduced $\chi^2$ (>5). However, for most of our Hot DOGs, the single \ser model still provides a reasonable fitting. Conservatively speaking, the \ser profile with $<n>=2.1$ can describe the distribution of most stars in Hot DOGs. Our result suggests that most of our Hot DOGs have an intermediate morphology (disk with a prominent bulge component). The intermediate morphology of Hot DOGs has been expected in the merger-driven scenario if Hot DOGs lie at the transient stage when major mergers with gas-rich disks are building galaxy bulge.
%Thus our result suggests that Hot DOGs may experience a morphological transform stage from disk-dominated to bulge-dominated.

Automatically morphological classification provides an independent check on the reliability of our visual classification. The $Gini$ and $M_{20}$ values of our Hot DOGs range from 0.47 to 0.79 with a median value 0.70, and from $-1.9$ to $-1.2$ with a median value of $-1.6$, respectively. We employ the $Gini/M_{20}$ parameters to locate major mergers using the merger criteria of Lotz et al. (2008). 13 out of 18 Hot DOGs fulfill the relation $Gini > -0.14M_{20} + 0.33$, indicating them as mergers. The high merger fraction ($\sim$72\%) derived here is well consistent with the result of our visual classification.

\begin{figure}
\plotone{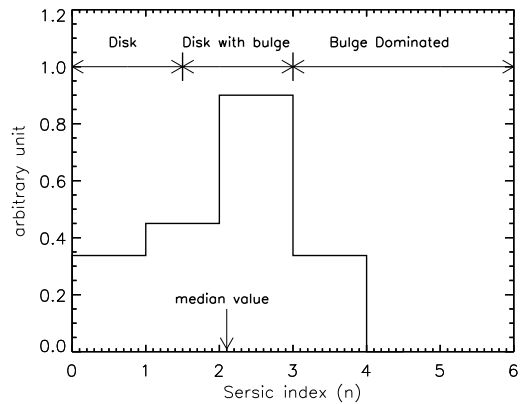}
\caption{The distribution of  \ser\ indices $n$ in the Hot DOG sample. The downward arrow marks the median value (2.1) of \ser\ indices $n$. }
\end{figure}

\section{Discussion and Summary}

In this letter, we use the high resolution \textit{HST} WFC3 near-IR imaging to derive the merger fraction in a subsample of 18 Hot DOGs by visual classification. We find that the merger fraction is high ($62\pm 14 \%$) for our Hot DOG sample. The high merger fraction in these most luminous, heavily-obscured AGNs are consistent with the predictions of hydrodynamical simulations \citep{hopkins2008} and the variability-based model \citep{hickox2014}.  We also derive the AGN bolometric luminosity by using IR SEDs decomposition. The heavily obscured AGNs in Hot DOGs are among the most luminous AGNs in the Universe ($\sim10^{14}L_\odot$). The cold dust emissions in Hot DOGs are also significant (see Table 2), indicating that there resides intense starburst in the hosts. Both intense AGN and starburst activities are likely triggered by mergers as suggested by the high merger fraction. However, the lack of a comparison sample may weaken the merger-driven explanation of Hot DOGs. Unfortunately, it is hard to select a suitable comparison sample, as both the stellar masses and black hole masses of Hot DOGs are not well known. The stellar mass upper bounds of Hot DOGs derived from SED modeling by \citet{assef2015a} are about $\sim10^{11.6}M_\odot$. Massive and inactive galaxies at $z>1$ show a relatively low fraction of strong distortion morphologies. For instance, only $<15\%$ galaxies in an inactive galaxy sample with the median stellar mass $\sim 5\times10^{10}M_\odot$ show major merger features \citep{fan2014}. This indirect evidence suggests that the high merger fraction in our Hot DOG sample is likely responsible for triggering the intense AGN activities.

According to fitting the surface brightness profiles, we find that the distribution of \ser indices of Hot DOGs peaks around 2, which suggests that most of Hot DOGs have intermediate morphologies. From the point of view of morphology, this work confirms our previous result (Fan et al. 2016a), which suggests that Hot DOGs may represent a transit phase during the evolution of massive galaxies, transforming from the dusty starburst dominated phase to the optically bright QSO phase.

We can find that all samples with high merger fraction are dust-reddened \citep{urrutia2008,glikman2015} and/or IR-luminous (this work) in Figure 2. While for the blue, luminous QSOs, no high merger fraction or no enhancing merging feature has been found. For instance, \citet{mechtley2015} recently studied an UV/optical-selected $z=2$ QSO sample with high super-massive black hole masses ($M_{BH}=10^9-10^{10}M_\odot$), which are comparable to the estimated SMBH masses of Hot DOGs \citep{assef2015a}. They found that the strong distortion fraction for QSO hosts is low and comparable to that of inactive galaxies. They therefore concluded that major mergers are not the primary triggering mechanism for AGN activity. The results between us look contradictory. However, the problem can be solved if considering that UV/optical-selected QSOs and IR-selected obscured QSOs represent distinct phases in the merger-driven evolutionary sequence \citep{sanders1988,alexander2012}.  In the merger-driven evolutionary model, gas-rich galaxy mergers fuel an intense starburst and a phase of obscured BH growth, followed by an unobscured phase after the gas is consumed or expelled from the galaxy by QSO feedback. Dusty starburst like SMGs and bump DOGs may represent the early stage of the mergers. They are expected to have a high merger fraction, which has already been observed (e.g., Bussmann et al. 2011). IR-luminous AGNs are in an intermediate stage of the mergers. Thus the strong merger features still present in their host galaxies. As a result, a high merger fraction will be derived according to morphological classification. While for unobscured QSOs whose hosts are found to be largely elliptical galaxies, they are in a late stage of mergers, just before the host galaxies settling into passive elliptical galaxies. We suggest that both high luminosity and obscuration should be essential for observing the high merger fraction in AGN hosts. We emphasize that this conclusion is only valid for radio-quiet AGNs while it may be another story for radio-loud AGNs \citep{chiaberge2015}. A similar evolutionary sequence has also been proposed by \citet{kocevski2015} involving X-ray selected AGN, infrared selected obscured AGN, dust-reddened quasars and unobscured AGN (see their Figure 10).

\acknowledgments

We thank the referee for the careful reading and the valuable comments that helped improving our paper.
This work is supported by the National Natural Science Foundation of China (NSFC, Nos. 11203023, 11303084, 11303002 and 11433005), the Fundamental Research Funds for the Central Universities (WK3440000001). LF acknowledges the support by Qilu Young Researcher Project of Shandong University. YH thanks the support from the Western Light Youth Project. GF acknowledges the support by the Yunnan Applied Basic Research Projects (2014FB155).

\textit{Facilities}: HST (WFC3)

\end{CJK*}

\end{document}